\documentclass[12pt]{article}
\pdfoutput=1
\usepackage{graphicx}
\usepackage[body={17.5cm, 22cm},right=2cm]{geometry}
\usepackage{amssymb}

\newcommand\ee{\end{equation}}
\newcommand\be{\begin{equation}}
\newcommand\eea{\end{eqnarray}}
\newcommand\bea{\begin{eqnarray}}

\def\r{\right}
\def\l{\left}
\def\eps{\varepsilon}
\def\beq{\begin{equation}}
\def\eeq{\end{equation}}

\def\op{{\cal O}}

\begin{document}
\setcounter{page}{0}
\thispagestyle{empty}

\begin{titlepage}

~\vspace{1cm}
\begin{center}

{\Large\bf (No) Eternal Inflation and Precision Higgs Physics}

\vspace{1cm}

{\large  \sc Nima Arkani-Hamed$^{a}$,  
Sergei Dubovsky$^{b,c}$, \\[.3cm]
Leonardo Senatore$^{b}$,  
Giovanni Villadoro$^{b}$}
\\[1cm]
{\normalsize { \sl $^{a}$School of Natural Sciences,\\
Institute for Advanced Study, \\Olden Lane, Princeton, NJ 08540, USA}}\\
\vspace{.2cm}
{\normalsize { \sl $^{b}$ Jefferson Physical Laboratory, \\ Harvard University, Cambridge, MA 02138, USA}}

\vspace{.2cm}
{\normalsize { \sl $^{c}$ Institute for Nuclear Research of the Russian Academy of Sciences, \\
        60th October Anniversary Prospect, 7a, 117312 Moscow, Russia}}
\end{center}
\vspace{.8cm}
\begin{abstract}

Even if nothing but a light Higgs is observed at the LHC, suggesting that the Standard Model is unmodified up to scales far above the weak scale, Higgs physics can yield surprises of fundamental significance for cosmology. As has long been known, the Standard Model vacuum may be metastable for low enough Higgs mass, but a specific value of the decay rate holds special significance: for a very narrow window of parameters, our Universe has not yet decayed but the current inflationary period can not be future eternal. Determining whether we are in this  window requires exquisite but achievable experimental precision, with a measurement of the Higgs mass to 0.1 GeV  at the LHC, the top mass to 60 MeV at a linear collider, as well as an improved determination of $\alpha_s$ by an order of magnitude on the lattice. If the parameters are observed to lie in this special range, particle physics will establish that the future of our Universe is a global big crunch, without harboring pockets of eternal inflation, strongly suggesting  that eternal inflation is censored by the fundamental theory. This conclusion could be drawn even more sharply if metastability with the appropriate decay rate is found in the MSSM, where the physics governing the instability can be directly probed at the TeV scale. 

\end{abstract}

\end{titlepage}

%%%%%%%%%%%%%%%%%%%%%%%%%%%%%%%%%

\section{Introduction}

The most remarkable recent discovery in fundamental physics is that the Universe is accelerating \cite{SSTC,SCPC,WMAP}. 
Will this acceleration last forever or eventually terminate? What will be the global picture of space-time, and is this a meaningful question? These are some of the most pressing theoretical issues of our time. The idea of eternal inflation \cite{Vilenkin:1983xq,Linde:1986fd} appears to be supported
by the existence of a landscape  of metastable vacua in the string theory 
\cite{Bousso:2000xa,Douglas:2006es}. It provides a plausible mechanism to populate the different vacua
making it possible to apply environmental arguments to explain the smallness of the cosmological
constant \cite{Weinberg:1987dv}. 
However,  the necessity to talk about causally disconnected space-time regions to describe eternal inflation
makes it extremely challenging to implement this picture in a full theory of quantum gravity.
At the moment we do not even know if there are any sharply-defined observables for such a theory to compute; if any exist, they likely reside at future boundaries, with no obvious connection to our own observations (see \cite{Witten:2001kn,Strominger:2001pn,Freivogel:2006xu,Susskind:2007pv} for some of attempts to make sense of this picture). This problem and the analogy of the emerging causal structure with that of the evaporating black hole, where a naive semiclassical logic fails by predicting the information loss, naturally suggests that semiclassical physics may also be misleading in this case,
and a qualitatively new approach is needed to describe the eternally inflating Universe (see, e.g.,
\cite{ArkaniHamed:2007ky} for a recent discussion). Clearly, in this situation any
experimental data shedding some light on the nature of the cosmological acceleration is extremely welcome.

At the classical level the choice is very simple. 
Either we live in a local minimum of energy and the cosmic acceleration
will last forever, or the acceleration is driven by the potential energy of the slowly rolling scalar field and may eventually
terminate. Things change qualitatively at the quantum level. In the slow roll case the Universe can still be eternally 
inflating if the acceleration parameter is sufficiently close to being constant~\cite{ArkaniHamed:2007ky},
\be
\label{HdH4}
{\dot H_\Lambda\over H_\Lambda^4}\lesssim M_{Pl}^{-2}\;,
\ee
where $H_\Lambda$ is the Hubble expansion rate corresponding to the current vacuum energy. Numerically, if the bound (\ref{HdH4}) does not hold  the current stage of the cosmological acceleration may last as long as $\sim
10^{130}$ years without being eternal. Current data show that the equation of state of the negative energy component is rather close
to that of the vacuum energy, $\dot{H}_\Lambda/H_\Lambda^2\lesssim{\rm few}\cdot 10^{-2}$. The bound (\ref{HdH4}) implies that the next
theoretically meaningful value for this ratio is at the level $H_\Lambda^2/M_{Pl}^2\sim 10^{-120}$. This indicates
that in the slow roll case the  chances to test  the eternal nature of the cosmological acceleration by directly measuring the time dependence of the acceleration parameter   $\dot{H}_\Lambda/H_\Lambda^2$ are not high in the next $10^{130}$ years or so. 

The situation is more interesting if we live in a local energy minimum. In the string landscape our positive energy vacuum 
is inevitably metastable towards decay into a lower energy state \cite{Coleman:1980aw}. Given the extreme smallness of the vacuum
energy this decay is likely to proceed into a rapidly crunching negative energy Universe. The decay goes through
the non-perturbative nucleation of  bubbles of the new vacuum that afterwards expand with the speed of light.
Whether or not the current stage of the cosmological acceleration is eternal depends on the rate $\Gamma$ of
the bubble nucleation. If the decay rate is fast enough
\be
\label{Gbound}
\Gamma\geq \Gamma_{\rm cr}={9\over 4\pi}H_\Lambda^4\;,
\ee
bubbles collide and eventually all the Universe suffers a transition into the new vacuum.
The current stage of inflation is not eternal in this case. On the other hand, for slower decay rates
the expansion of the Universe wins and bubbles never fill the whole inflating volume.
Of course, bubble walls propagate with the speed of light so we have no chances of observing bubbles
without being immediately killed by a domain wall. However, in principle,
we can establish whether the bound (\ref{Gbound}) holds by performing high precision
measurements of the underlying particle physics parameters, so that  it becomes possible to identify a nearby
negative energy minimum and to calculate a lower bound for the decay rate with high enough precision. We say `a lower bound' because there are always potential instabilities coming from high energy physics we cannot control, but it suffices to find fast enough infrared calculable instabilities to see that inflation cannot be eternal. 

High precision measurements are needed since, given that the Universe did not decay up to the present we know that the decay rate $\Gamma$ cannot be 
significantly faster than the lower bound (\ref{Gbound}). Imposing that the probability ($p$) not to decay up to now is larger than, say, 5\% ($2\sigma$) gives $\Gamma\leq 52.3\,\Gamma_{\rm cr}$, for generic $p$ the rate will be in the window
\beq \label{eq:decayreg}
\Gamma_{\rm cr}\leq\Gamma\leq 17.45\, \log{(p^{-1})}\, \Gamma_{\rm cr}\,.
\eeq
Given that $\Gamma$ depends exponentially on the underlying particle physics parameters, we will see that  even a change in the value of $\Gamma$ by a factor of 50 corresponds to a
tiny change of particle masses and coupling constants.

In this paper we argue that, with a certain 
amount of luck,   challenging but feasible measurements at the TeV scale may establish the existence of a fast enough instability, such that the bound (\ref{Gbound}) holds. Our main focus will be a minimal scenario of this kind, which is realized if there is no
new physics up to  very high energies of order the GUT scale. As reviewed in section~\ref{sec:SM},
in this case if the Higgs mass is light enough, the Standard Model (SM) potential develops a negative
energy minimum at large values of the Higgs field. We show then that one can verify 
whether the bound (\ref{Gbound}) holds for this particular decay channel by performing high precision
measurements of the Higgs mass (with uncertainty $\Delta m_H\sim 0.2$~GeV), the top mass
($\Delta m_t\sim 60$~MeV) and the strong coupling constant ($\Delta\alpha_s/\alpha_s\sim 10^{-3}$).
These numbers are achievable with the future data from the LHC and a linear collider. On the theoretical side one will need to
improve the existing calculations of the decay rate by including at least one extra electroweak and two
strong loops. 

Of course, we will never be able to directly verify that the Standard Model holds up to the GUT scale. However, the window for the SM parameters allowing the decay rate to be fast enough, such that the
current cosmological acceleration is not eternal, is extremely narrow. Consequently, if no new physics
is found at the LHC and the parameters turn out to be in that window, it is reasonable to take
this scenario seriously and to try to find an underlying reason why Nature worked so hard to avoid eternal inflation. We outline reasons why Nature might want to censor eternal inflation  in section~\ref{sec:no}. We stress that we do not necessarily find these arguments plausible---oNe of the AutHors doesn't believe them---they certainly require what looks like a conspiratorial restriction on particle physics models, (although milder examples of surprising gravitational limitations on effective field theories have been seen in other contexts \cite{Vafa:2005ui,ArkaniHamed:2006dz,Ooguri:2006in}).  It seems much more reasonable that eternal inflation {\it does} occur, and that we have to learn how to properly deal with it.  Nonetheless the counter arguments are not unreasonable, and it is interesting that entirely feasible experimental observations could strongly push us to take them seriously.

These arguments {\it predict} that the vacuum decay rate is within the region of eq.~(\ref{eq:decayreg}).
{\it A priori}, there is no reason to expect this region to be so narrow.
However it  is very narrow in the real world. 
This is nothing but a reflection of the famous cosmic coincidence problem (i.e. why $T_{\rm U}\sim H_\Lambda^{-1}$). 
At the moment, the best solution to this problem is provided by the
Weinberg's argument~\cite{Weinberg:1987dv}. However, we would like to stress that the fact that the
non-eternal inflation bound predicts a very narrow range in the Standard Model parameters 
is independent of whether the Weinberg's explanation is the correct one and 
is just a consequence of experimental data. On a similar note, observing the Standard Model parameters 
in this tiny window would  neither support nor disfavor Weinberg's argument.
  
The situation can be even more interesting if new physics is discovered at the LHC, as in this case
a negative energy minimum could be found at directly accessible energy scales. In section~\ref{sec:MSSM} we briefly review
one scenario of this kind, which is supersymmetry with large soft-breaking trilinear couplings.  
We conclude by summarizing our results and outlining future directions.

\section{The Standard Model Analysis}
\label{sec:SM}
\begin{figure}[t!]
\begin{center}\includegraphics[width=0.6\textwidth]{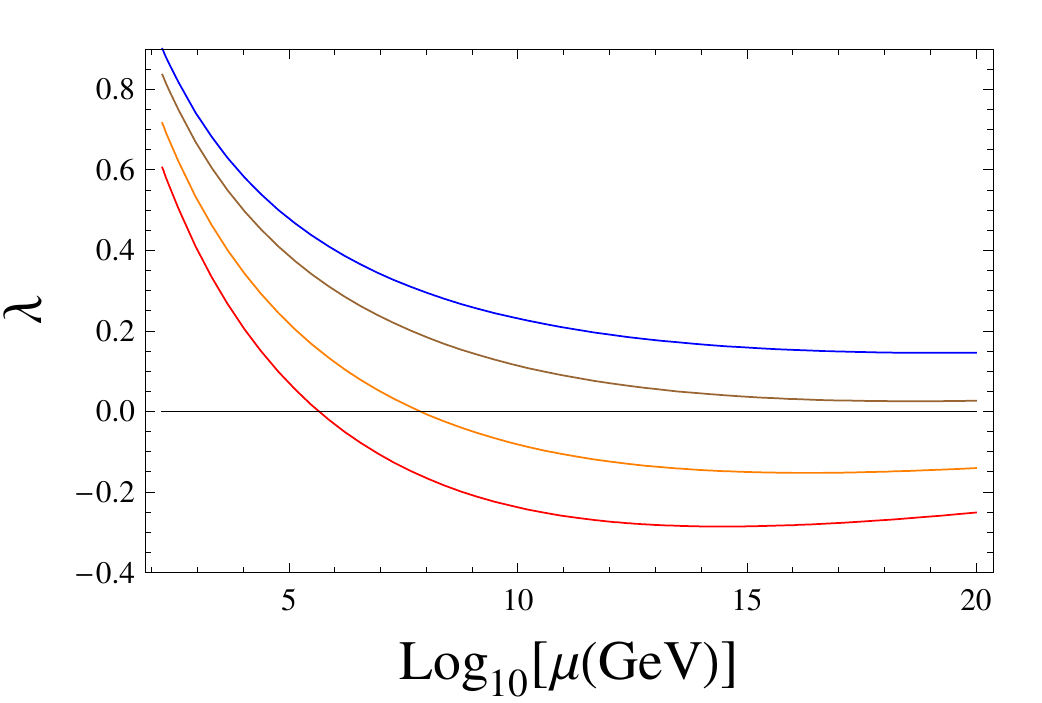}\end{center}
\caption{Running of the Higgs quartic coupling $\lambda(\mu)$ for $m_t=170.9$~GeV for different
Higgs masses (from below) $m_H=110$, $120$, $130$ and $135$~GeV (see appendix for formul\ae\ and notations). 
For low masses $\lambda$ turn negative at high scales $\mu$ and the SM vacuum become unstable.
\label{fig:lambda}}
\end{figure}

In the case the LHC will discover nothing else but the Higgs up to the TeV scale, 
it is reasonable to assume (as we will do in this section) that the Standard Model may hold up to very large scales,
say the seesaw, the GUT or the Plank scales. In this scenario it is well known 
\cite{Krive:1976sg}-\cite{Espinosa:2007qp} that the Higgs
potential may develop an instability that makes our vacuum decay. This is due to the fact
that, depending on the physical Higgs and top masses, the running may turn the
quartic coupling, and thus the potential, negative at high scales (see fig.~\ref{fig:lambda}). 
Requiring our vacuum to be sufficiently long lived with respect to the present age 
of the Universe ($T_{\rm U}$) a lower bound on the Higgs mass has been derived~\cite{IRS}, 
giving\footnote{Using the updated values for the top mass and the strong coupling constant
given below.
%$m_t=172.6\pm0.8_{\rm stat}\pm 1.1_{\rm syst}\simeq172.6\pm1.4$~GeV \cite{CDF},
%$\alpha_s=0.1176\pm 0.0020$ \cite{PDG}.} \cite{IRS}
}
\beq m_H\gtrsim 111 \ {\rm GeV}\,. \label{eq:boundIRS}
\eeq 
%\beq \label{eq:boundIRS}
%m_H({\rm GeV})>117+2.9\l(m_t({\rm GeV})-175\pm2\r)-2.5\l(\frac{\alpha_s(m_Z)-0.118}{0.002}\r)+
%0.1\log\l(\frac{T_{\rm U}}{10^{10}{\rm yr}}\r)\,,
%\eeq
%where $m_t$ is the top pole mass and $\alpha_s(m_Z)$ is the $\overline {\rm MS}$ strong coupling constant at
%the $Z^0$ mass $m_Z$. 
This decay channel disappears when
$m_H\gtrsim 129$~GeV.
% (note that this value is somewhat higher than the one shown in fig.~\ref{fig:lambda}
%due to the higher order effects included in the plot, see below).
%\beq
%m_H({\rm GeV})>133+2.0\l(m_t({\rm GeV})-175\pm2\r)-1.6\l(\frac{\alpha_s(m_Z)-0.118}{0.002}\r)\,.
%\eeq

If we assume that the current observed acceleration of the Universe 
is due to the vacuum energy (as opposed to some quintessence field) we are led to the following fundamental question: 
\emph{will our Universe eternally inflate or not?} As we will show below there is a critical value for the Higgs mass
such that if the Higgs is lighter, we will be able to say that our Universe will not inflate forever.
%Let us explain how this is possible. %As pointed out in \cite{G,GW}, 
Inflation in a false vacuum (with a positive cosmological constant $\Lambda=3 H^2_\Lambda /(8 \pi G_N)$\,) may or may not be eternal depending 
on whether the decay rate $\Gamma$ is smaller or larger than a critical value.
Indeed the inflating volume as a function of time is
\beq \label{eq:inflvol}
V_{\rm infl} = V_0 \, e^{3 H_\Lambda t} e^{-\Gamma\, \widehat{\rm Vol}_4(t)},
\eeq
where $\widehat{\rm Vol}_4(t)=(4\pi/3 H_\Lambda^3)\,t$ is the 4-volume spanned by the past light-cone of a comoving point after a time $t$~\cite{GW},
the first factor is due to the de-Sitter expansion while the second represents the exponential decay due to bubble nucleation.
When $\eps\equiv\Gamma/H_\Lambda^4<9/4\pi$ the expansion rate of the Universe 
wins against the decay rate, the inflationary volume expands exponentially and the 
Universe eternally inflates. In the opposite case  bubbles percolate and inflation ends globally in a finite 
time.
%\footnote{In eq.~(\ref{eq:inflvol}) we neglected the corrections from bubble overlap, for this reason the value for $\eps$ could
%deviate from $9/4\pi$; we expect the deviation to be at most of order one, for this reason in our results we will keep
%the dependence on $\eps$ explicit.}. 

Notice that just requiring that bubbles percolate is not enough to guarantee that inflation ends globally;
indeed in \cite{GW} the critical value for percolation has been identified to lie in the interval 
\be
\label{truepercolation}
1.1 \cdot 10^{-6} < \eps < \frac{9}{4 \pi} n_c= 0.24\,,
\ee
which is smaller than $9/4 \pi$ (here $n_c=0.34$ is the critical ratio between the volume in the bubbles and the total volume). In what follows, we are being somewhat sloppy in the terminology and 
by percolation transition we refer to the transition to the non-eternal regime, $\eps=9/4\pi$.

%
%The precise value for $\eps$ is still unknown, however the range of the possible values has been identified in~\cite{GW}:
%\beq
%1.1 \cdot 10^{-6} < \eps < 0.24 \,.
%\eeq
%Although this interval of values seems to be quite large as we will see 
%it will induce a very small uncertainty on the bound for the Higgs mass;
%further $\eps$ is expected to lie close to the upper bound~\cite{GW}.

In order to translate the bound $\Gamma/H_\Lambda^4>\eps$ into a bound for the Higgs mass we need
to know how $\Gamma$ depends on $m_H$. The expression for the decay rate 
per unit space-time volume is \cite{Kobzarev:1974cp,C,CC}
\beq \label{eq:gamma}
\Gamma=\frac{1}{V_4} e^{-S_E}\,,
\eeq
where $S_E$ is the euclidean action calculated on the bounce solution, $V_4=R^4$ and $R$ is the size of the bubble
that maximizes the rate~\cite{IRS}. 
At one loop $S_E$ reads
\beq
\label{ecaction}
S_E=\frac{16 \pi^2}{|\lambda(\mu)|}+\Delta S(\mu R)\,,
\eeq
where $\lambda(\mu)$ is the Higgs quartic coupling at the scale $\mu\simeq R^{-1}$ and $\Delta S(\mu R)$ represents 
the finite and the not log-enhanced one-loop corrections, which have been computed in \cite{IRS}.
This expression can be improved by including the two-loop log-enhanced corrections through
resumming up to two loops the RGE evolution of the couplings~\cite{LSZ,FJSE} from the weak scale to the scale of the bubble $R^{-1}$.
Requiring that $\Gamma>H_\Lambda^4 \eps$ gives an upper bound on $S_E$; in turn this gives a
lower bound to $|\lambda(\mu\simeq 1/R)|$. Since $\lambda(\mu\simeq 1/R)$ is negative
this finally gives an upper bound to the quartic coupling at the weak scale, thus to the Higgs mass.

Before entering the details of the calculation of $\Gamma$, we can anticipate some features of the result.
The upper bound from not being eternally inflating will be very close to the 
stability lower bound (\ref{eq:boundIRS}) because the dependence on $T_U$ of the latter is logarithmic 
and, as it happens in our Universe, $T_{\rm U}\sim H_\Lambda^{-1}$. 
The precision required in order to determine whether we are not eternally inflating will be  
set by the difference between the  (would be) observed Higgs mass and the one saturating the no eternal inflation bound. We can 
estimate an upper bound to this difference by imposing the probability that our Universe has not decayed yet to be 
larger than $p$, 
i.e.
%\footnote{Here ${\rm Vol}_4(T_U)\simeq 0.08\cdot H_\Lambda^{-4}$ is the space-time volume in the past lightcone of an observer at time $T_U$.}
\beq
\label{lifetime}
e^{-{\rm Vol}_4(T_U)\,\Gamma} > p\,,
\eeq
and plugging ${\rm Vol}_4(T_U)\simeq 0.08\cdot H_\Lambda^{-4}$ for the space-time volume in the past lightcone of an observer at time $T_U$. 
As we will see momentarily (see eq.~(\ref{eq:res}) below), in this way one  obtains a lower bound on the Higgs mass that differs from the percolation bound by
\beq
\Delta m_H({\rm GeV})=0.05 \log \l (\frac{9}{4\pi}\frac{0.08}{\log p^{-1}} \r)\,,
\eeq
which gives $0.2$~GeV and $0.15$~GeV for $p=0.05$~($\sim2\sigma$) and $p=0.32$~($\sim1\sigma$) respectively.
%\footnote{
%The reason why this value is somewhat smaller than the one in eq.~(\ref{eq:boundIRS}) is mainly due
%to the higher order corrections we included with respect to \cite{IRS}.} (\ref{eq:res}).
%Note that this value is somewhat lower than the one in eq.~(\ref{eq:boundIRS}). The difference is mainly due
%to the higher order corrections we included with respect to \cite{IRS}.
This means that if the Higgs potential prevents the Universe from being eternally  inflating, 
the measured Higgs mass must lie in the tiny window within $0.2$~GeV below
the percolation bound. 

For the same reason the radius of the bubble will be close to
that found in \cite{IRS}, i.e. $R^{-1}\sim 10^{17}$~GeV. The reason for such a small scale is that
an optimal bubble radius is set as a result of a competition of different logarithmic effects---the top contribution to the quartic coupling running making it negative, and the gauge contributions to the running pushing  the quartic coupling up.

Given that  high precision is needed and since we have to run for a huge interval of scales
we want to compute the rate with the best possible accuracy. 
In order to perform the running of $\lambda(\mu)$, as done in \cite{FJSE} 
we integrated the two-loop RGE equations of the three gauge couplings, 
the top Yukawa and the quartic couplings. This corresponds to including the contributions 
at the following orders: $\op(\alpha_W^2)$, $\op(\alpha_s^2)$ and $\op(\alpha_W \alpha_s)$,
where $\alpha_W$ stays for both the electroweak and the top Yukawa couplings. We also included 
three-loop $\op(\alpha_s^3)$ contributions (for the top Yukawa~\cite{T,L2} and the strong coupling constant~\cite{TVZ,LV}) 
to the running because they are known and are comparable to those of order $\op(\alpha_W \alpha_s)$ (since $\op(\alpha_W)\sim\op(\alpha_s^2)$). At this order we also need the 
initial condition for the top Yukawa and quartic coupling matched at one-loop for 
the Higgs pole mass~\cite{SZ} and up to one-loop weak~\cite{HK} and two-loop strong~\cite{GBGS,BGS,FJTV} 
for the top pole mass. $\op(\alpha_s^4)$ corrections to the $\alpha_s$ running~\cite{vRVL,Czakon:2004bu}, 
three-loop strong~\cite{MR} and mixed two-loop strong/weak~\cite{JK} contributions for the
top mass are also known and we used them to estimate higher order corrections. 

Using the preliminary value for the top mass in \cite{CDF} $m_t=172.6\pm0.8_{\rm stat}\pm1.1_{\rm syst}\simeq172.6\pm1.4$~GeV,
and the current value~\cite{PDG} of the strong coupling $\alpha_s=0.1176\pm 0.0020$  we get the no-eternal inflation bound
(see the appendix for the details of the calculation)
\beq \label{eq:res}
\hspace{0pt} m_H({\rm GeV})<109.1+4.4\times \frac{m_t({\rm GeV})-172.6}{1.4}-2.5\times \frac{\alpha_s(m_Z)-0.1176}{0.0020}
	+0.05\log\l(\frac{9}{4\pi \eps}\r)\pm3_{\rm th}\,,
\eeq
with bubbles of size $R^{-1}\simeq 10^{16}$~GeV for Higgs masses saturating the bound.
The theoretical error in eq.~(\ref{eq:res}) is only indicative: we estimated it by including the effects of the known 
higher order corrections in the matching of the top mass, the running of $\alpha_s$ and by varying the matching 
scale\footnote{For the central value of the bound in eq.~(\ref{eq:res}) we matched the top Yukawa at $\mu=m_t$, 
the Higgs quartic coupling at $\mu=m_H$ and the gauge couplings at $\mu=m_Z$, see the appendix for more details.} 
$\mu$ between $m_Z$ and $v=246.22$~GeV. 
All these methods give a shift to the bound (\ref{eq:res}) of about 3~GeV or less. As usual the estimation
of the theoretical errors is very uncertain and should be taken at the level of order of magnitude. 
\begin{figure}[t!]
\begin{center}
\begin{tabular}{c}\includegraphics[width=0.55\textwidth]{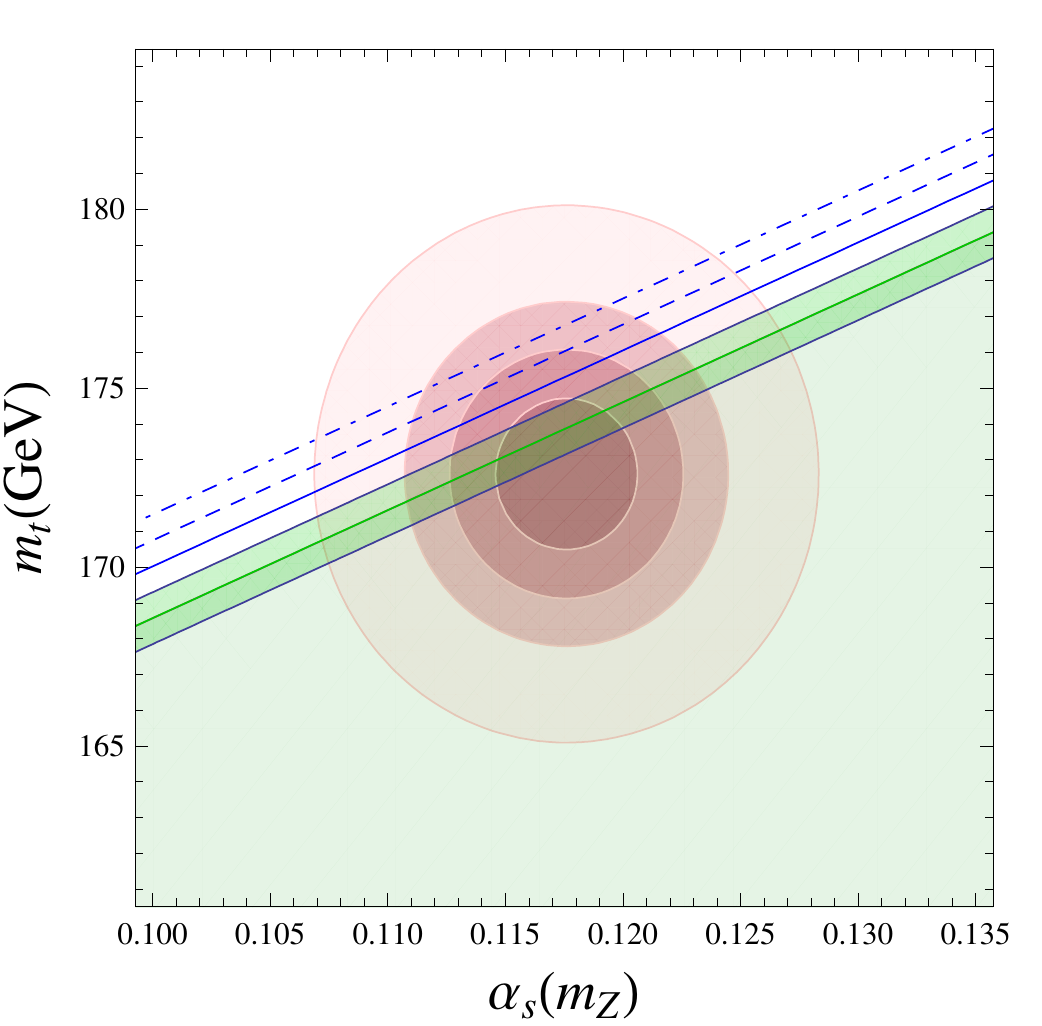} \\ ~ \end{tabular}
\begin{tabular}{c}\includegraphics[width=0.25\textwidth]{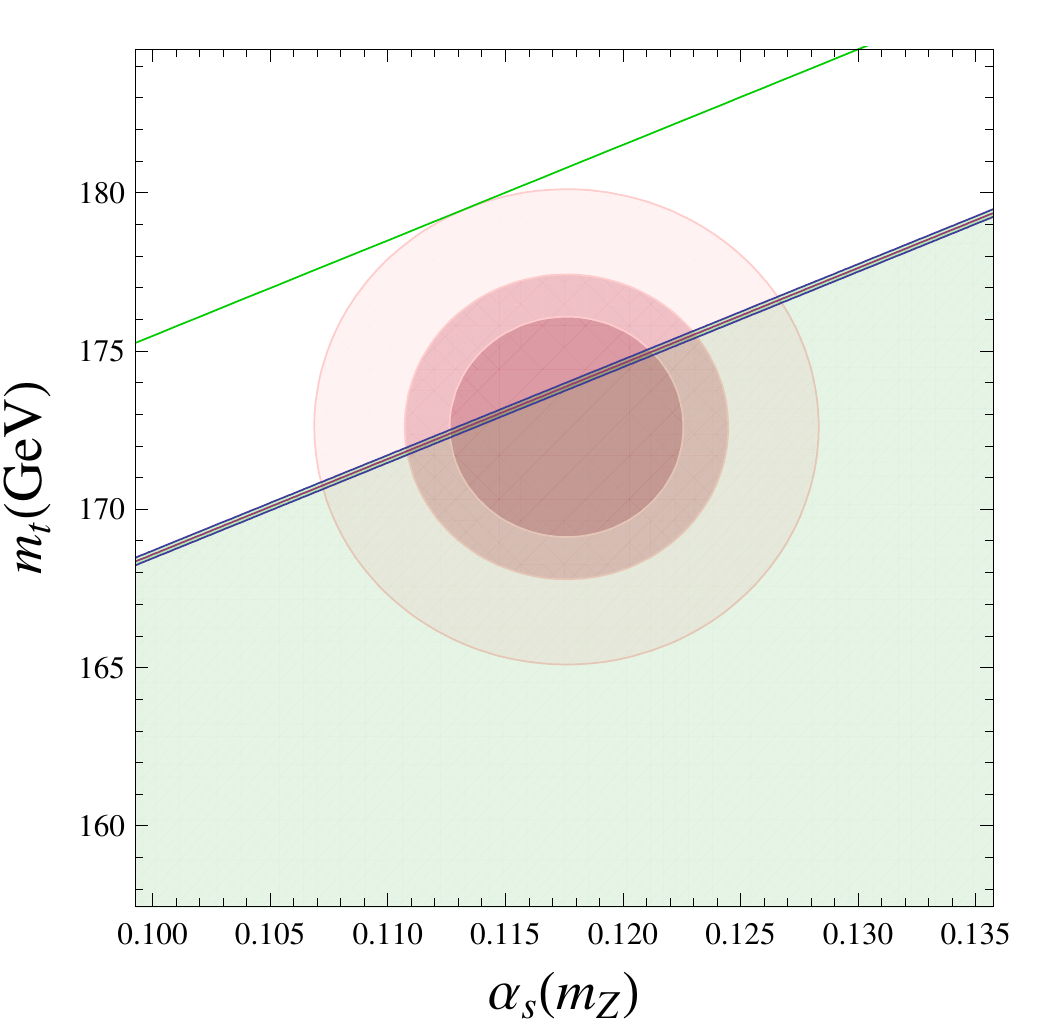} \\
\includegraphics[width=0.25\textwidth]{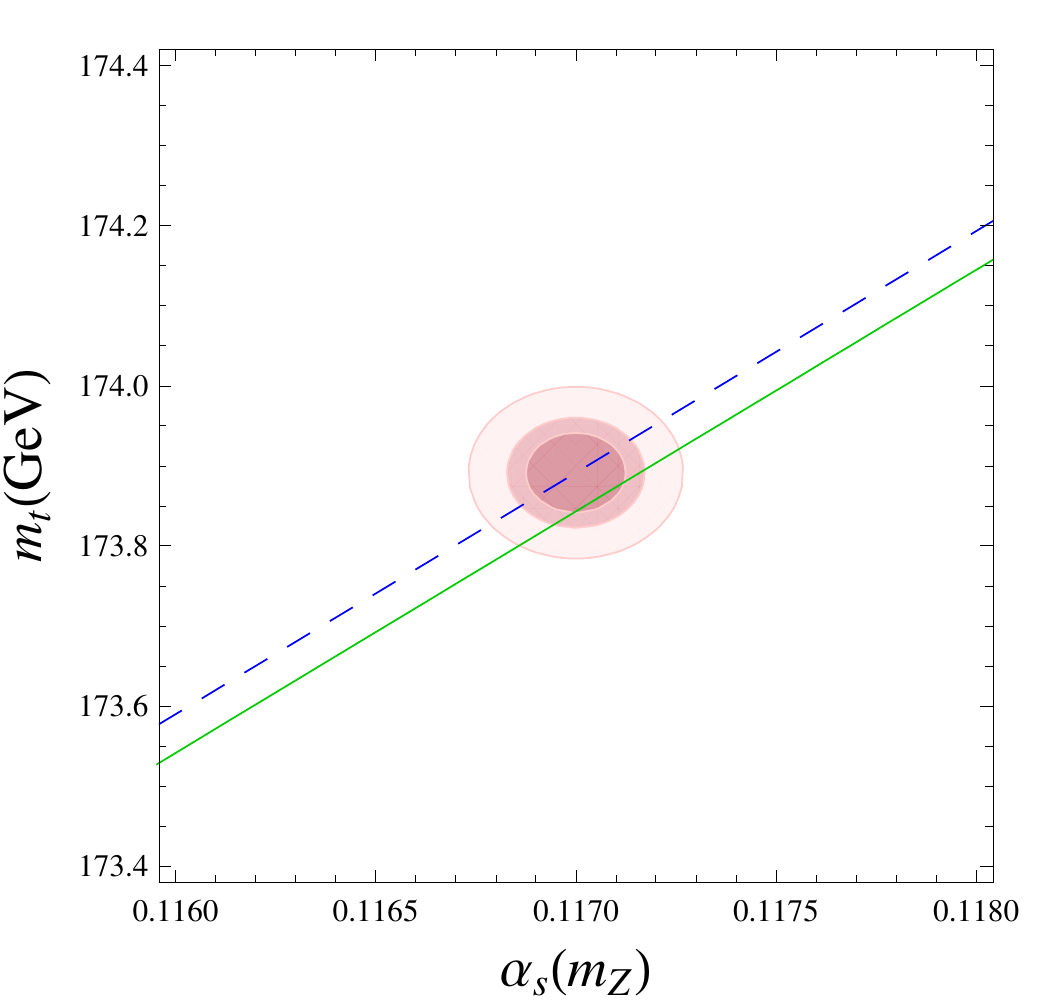} \\ ~ \end{tabular}
\end{center} \vspace{-1.1cm}
\caption{\small \emph{Left:} measured $m_t(GeV)$ and $\alpha_s(m_Z)$ at 1, 2, 3 and 5$\sigma$ (ellipses)
\emph{vs} the percolation bound for different Higgs masses (blue lines). The line in the middle of the
shaded strip corresponds to the LEP2 bound $m_H=114.4$~GeV (the excluded experimental region is green shaded), 
all the other lines are at step of 3 GeV in the Higgs mass. The shaded strip corresponds to the $\pm 3$~GeV 
theoretical error on the bound. This shows that, with the current experimental and theoretical errors,
non-eternal inflation is excluded at less than $1\sigma$ level, i.e. it is compatible with current observations.
\emph{Right up:} Rule out scenario: all the values stay the same and an heavy Higgs mass is observed ($m_H\simeq 145$~GeV), 
eternal inflation is not excluded by the bound; \emph{Right down:} Rule in scenario: both the experimental
and the theoretical errors have been substantially decreased and the central values moved such that
the bound on the Higgs mass (dashed blue line) ($m_H=115.2$~GeV in the figure) is above the experimental value (green line,
$m_H=115$~GeV in this case) by $\sim2\sigma$; the Universe does not experience eternal inflation. \label{fig:topalpha}}
\end{figure}

Since the direct search for the Higgs at LEP2 \cite{LEP2} gives a lower bound $m_H>114.4$~GeV at 95\% CL, which
is within $1\sigma$ from (\ref{eq:res}), the discovery of a light Higgs may signal 
that we are not eternally inflating. In this case in order to be sure (see fig.~\ref{fig:topalpha})  one should reduce
the theoretical and experimental uncertainties and check whether the Higgs mass is above or below the bound (\ref{eq:res}).
After LHC the scenario will be reversed; while the Higgs mass will be known with high accuracy, the precision
on the top mass will not improve much, at this point it will become more useful to rewrite the bound (\ref{eq:res})
as a bound on the top mass:
\beq
m_t({\rm GeV})>174.4+0.3\times (m_H({\rm GeV})-115)+0.8\times \frac{\alpha_s(m_Z)-0.1176}{0.0020}\pm1_{\rm th}\,.
\eeq
Note that in this case also the size of the bubble shrinks. For the choice of the parameters such that
the percolation bound is saturated at $m_H\sim 115$~GeV, the size is   $R^{-1}\sim 10^{17}$~GeV.

On the contrary if the Higgs turns out to be heavier than the bound (\ref{eq:res}) 
we will not be able to come to a definite conclusion on the fate of the Universe. Indeed, the rate $\Gamma$
can always receive contributions from unknown physics (even at the Planck scale)
that can make the decay rate large enough 
to avoid eternal inflation. Therefore our bound, together with the assumption that the running is not 
affected by new physics up to large enough scales, allows us to determine the future evolution of the Universe
only in the case the Higgs mass satisfies eq.~(\ref{eq:res}).

Using the bound (\ref{eq:res}) it is straightforward to see that the range of $\Gamma$ where we are not eternally inflating and 
 the probability that our Universe has not decayed yet is larger than 5\%, correspond to a tiny window for the Higgs mass within $0.2$~GeV below
the bound (\ref{eq:res}). In order to verify the no eternal inflation hypothesis all the uncertainties in eq.~(\ref{eq:res}),
i.e. those in  the top and the Higgs masses, $\alpha_s$ and the theoretical uncertainties must be reduced at this level.
This is a challenging task both experimentally and theoretically.

On the experimental side
one would need to reduce the error on the top mass by a factor of $\sim20$, the one on $\alpha_s$ by
one order of magnitude and to measure the Higgs mass with an error less than a fifth of GeV.
The latter should be within the reach of LHC~\cite{TDRs}. It should also be able to increase somewhat
the precision on $m_t$, although the required precision on the top mass will be probably
achieved only at a linear collider. On the other hand, the most precise determination of $\alpha_s$ at the moment comes 
from lattice simulations and the situation does not seem to change with the forthcoming experiments.
Hopefully numerical calculations will be able to increase their power (at the same time reducing  systematic
uncertainties) by an order of magnitude on the decades time-scale. 

On the theoretical side, both the running 
and the matching conditions for gauge couplings and masses should be improved at 
least by one extra electroweak loop and two strong loops. 
This also implies an extra loop in the calculation of the bouncing solution
and the corrections to the corresponding action in the decay rate expression.
% Finally, the error associated 
%with corrections to the  coefficient $\eps$ is limited because of the
%very weak (logarithmic) dependence on $\eps$ in eq.(\ref{eq:res}). 

Our working assumption so far was  that the physics up to the scale of the bounce,  $R^{-1}\simeq 10^{17}$~GeV, is exactly that of the Standard Model.  However, there is one set of corrections that we definitely cannot ignore---those coming from gravity. Given the relatively small size of the bounce and  the high precision we need, one may worry that gravitational corrections can introduce a substantial uncertainty in Higgs mass bound  (\ref{eq:res}). Fortunately, this is likely to be not the case in the most interesting
range of parameters $m_H\simeq 115$~GeV.

Indeed, in general, there are two sorts of gravitational corrections. First, there are corrections  that can be reliably calculated within effective field theory. The leading correction of this kind is due to the gravitational
contribution to the tree level bounce action. A recent calculation~\cite{Isidori:2007vm}
provides the following handy analytic approximation for this correction
\be
\label{treegrav}
\Delta S_{grav}\approx {1024\pi^3\over 5 (RM_{Pl}\lambda)^2}\;.
\ee
This contribution is straightforward to include in our analysis, and at the required level of precision it is negligible for Higgs masses up to $m_H\simeq 120$~GeV
(with the appropriately shifted central values of $m_t$ and $\alpha_s$ such that this 
Higgs mass saturates the percolation bound). For the values of the parameters such that the percolation bound is saturated at $m_H\simeq 120$~GeV, the correction in (\ref{treegrav}) shifts $m_H$ by $0.1$~GeV, while for smaller $m_H$ the correction rapidly goes to zero.

The log-divergent graviton loop contribution is also 
calculable and is further suppressed with respect to the tree level correction (\ref{treegrav}).
The incalculable corrections arise due to our ignorance about the full theory of quantum gravity, and correspond to the power divergent loop contributions in the effective theory.
They can be conveniently parameterized by the higher-dimensional operators of the form
\[
\Delta {\cal L}={c\over M_{Pl}^2}{\phi_h^6\over 6!}+\dots
\]
in the effective action, where $\phi_h$ is the Higgs field and $c$ is a dimensionless coefficient.  The effect of these operators on the bounce action (\ref{ecaction})  can be approximated as
\be
\label{lshift}
\Delta S'_{grav}\sim {cR^4\over M_{Pl}^2}\phi_{b}^6\sim{c\over \lambda^3 (R
 M_{Pl})^2}\;,
\ee
where at the last step we plugged in the value of the Higgs field in the center of the bounce
$\phi_{b}\sim\lambda^{-1/2}R^{-1}$. 

The presence of such a correction is equivalent to the
additional uncertainty in the determination of the quartic coupling at the bounce scale of order
\be
\label{dll}
{\delta\lambda\over\lambda}\simeq {c\over \lambda^2 (R
 M_{Pl})^2}\;.
\ee
Note now that the uncertainty in the quartic coupling  $\lambda$ at high scales can be related to 
the shift in the prediction for the Higgs mass as
\begin{equation}
\frac{\delta m_H}{m_H}=\frac{1}{2}\frac{\delta \lambda}{\lambda}(\mu=m_H)\sim \frac{\delta \lambda}{\lambda}(\mu=1/R)\ . 
\end{equation}
Since we need a precision of order $\lesssim 0.2$~GeV on the Higgs mass, we need $\delta\lambda/\lambda(\mu=1/R)\lesssim 10^{-3}$.  

At the effective field theory level there is no reason to expect this correction to be large unless
the bounce energy-density, which is of order $\lambda^{-1}R^{-4}$, reaches the Planck scale.
This expectation is in agreement with (\ref{dll}) if $c\geq{\cal O}(\lambda^{3/2})$ and, indeed,
diagrams present in the effective field theory generate operators whose contributions are at most 
of order of the one in (\ref{lshift}) with $c={\cal O}(\lambda^{3/2})$.
This kind of corrections are always smaller than the calculable one (\ref{treegrav}) and
are never important in the interesting range of parameters.

Note, however, that in quantum gravity, in particular in string theory, one may expect larger
values of $c$. Indeed, at the perturbative level the reason why $c$ is small for small $\lambda$ is that
it is protected by the shift symmetry that gets restored at $\lambda=0$. However, in quantum gravity 
global symmetries are always broken and,  in principle, one may have even $c\simeq 1$. For instance,
in the context of string theory, scalar fields usually have some geometrical meaning, so that the size
of higher order corrections is typically controlled by fields vev's in Planck or string
units rather than just by the energy density. 
%Still, given the smallness of the quartic coupling
%it is natural to expect that parameters controlling  the size of the higher order operators are also suppressed by the weak coupling parameter.

In the worst case scenario $c\simeq 1$ the incalculable correction (\ref{lshift}) is enhanced by a factor
$1/\lambda$ with respect to the tree level contribution (\ref{treegrav}).
As a result it gives a shift of $0.1$~GeV if the parameters are such that the percolation bound is saturated at $m_H\simeq117$~GeV,
as before, for smaller $m_H$ the shift decays quickly.

\begin{figure}[t!]
\begin{center} \includegraphics[height=7cm]{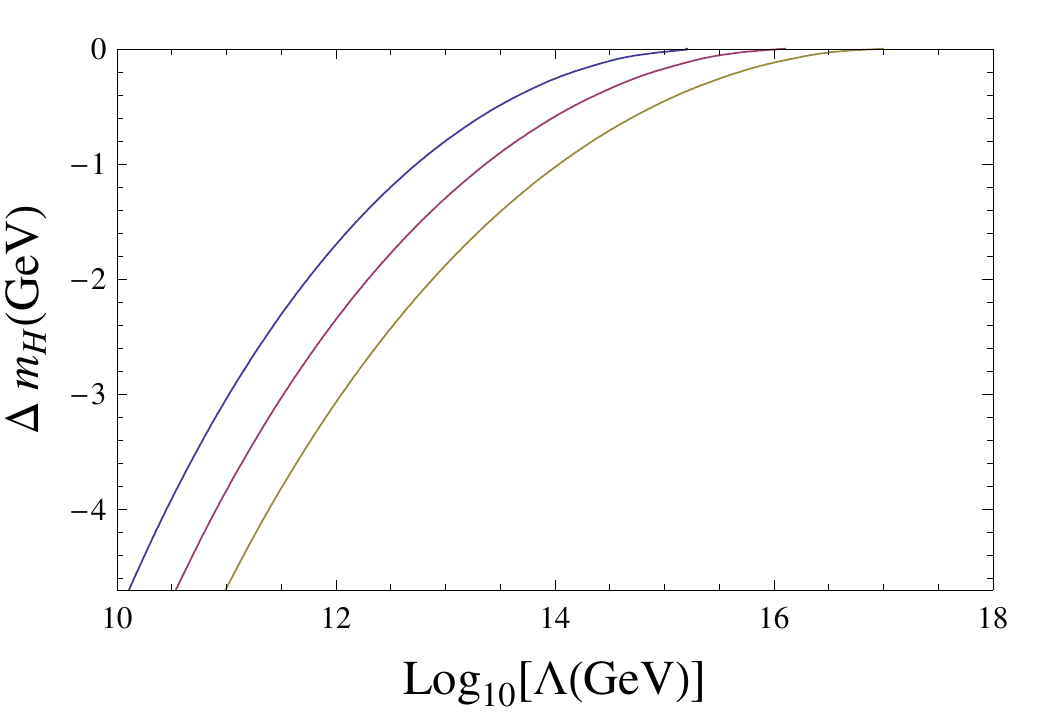} 
\end{center}
\caption{\small Decrease of the upper bound on the Higgs mass in eq.(\ref{eq:res}) from lowering the cutoff $\Lambda$ for various values of the top mass (from the left $m_t=171$, $173$ and $175$~GeV).} 
%Right: Value of the cutoff $\Lambda$ for a given value of the top mass $m_t$ and of the upper bound of the Higgs mass $m_H$ in eq.(\ref{eq:res}).}
\label{fig: cutoff_dependence}
\end{figure}

We see that even under the most pessimistic assumptions gravitational corrections are only marginally 
important in the interesting parameter range. Still, we can take a conservative attitude and lower the maximum scale at which the calculation should be trusted. As we lower the maximum UV scale, the decay rate decreases, and therefore the percolation bound on the Higgs mass (\ref{eq:res})  goes down.  In fig.~(\ref{fig: cutoff_dependence}) we show how the bound on the Higgs mass decreases as we decrease the cutoff $\Lambda$ of our calculation for various top masses. We see that imposing the cutoff to be as small as $\sim 10^{16}$~GeV, so that even the worst case scenario gravitational corrections can be safely ignored, amounts to decreasing the upper bound on the Higgs mass only by a tiny amount. We further see that the bound on the Higgs mass does not get very much decreased (only 
by a few GeV) by imposing the cutoff to be as small as  $10^{12}$~GeV, a value which allows to accommodate GUT and seesaw.

Related to this point, one may wonder what would be the implications if in this scenario the Higgs happened to be much lighter (of the order of a few GeV) than the bound of (\ref{eq:res}).
Given that the percolation bound is very close to the life-time bound (\ref{lifetime}), in order
to avoid the conclusion that our probability to survive up to now is ridiculously small, as it is very well known, one is
forced to conclude that the scale of new physics is rather low, see e.g.~\cite{CEQ,SW}.
With this in mind, fig.~(\ref{fig: cutoff_dependence}) would tell us the scale of new physics.
% (right) shows the value of the cutoff to be imposed for a given upper bound to the Higgs mass $m_H$ from eq.(\ref{eq:res}) and top mass $m_t$. Given the vicinity of the bounds from not-eternally inflating and from the lifetime of the Universe, the cutoff one would infer from imposing a high enough probability of survival up to the present would give a very similar result for the cutoff.
%

%Finally, it is worth noting, that we definitely know that there is some physics beyond the Standard
%Model---one providing a source for the neutrino masses. There are two main options to give mass to the neutrinos and explain their oscillations. One possibility is that there are three right handed neutrinos and they have Dirac masses. In this case the corresponding Yukawa couplings are tiny and, therefore, irrelevant for our calculation. A second possibility, the so called seesaw mechanism, is based upon giving a Majorana mass to the right-handed neutrinos of the order $\sim 10^{16}$~GeV, assuming the Yukawa coupling to be of order one. In this case, the effect of the Yukawa on the running will be negligible, since the neutrinos enter only around the scale of the bounce. On the other hand, finite correction to the bounce solution in this case may be marginally relevant shifting the bounce action at the level $\Delta S\sim 1$. 

\subsection{Vacuum decay in the MSSM}
\label{sec:MSSM}
Although it is possible that the Standard Model is valid up to high scales, it
is not natural from an effective field theory point of view. It is highly
expected that new physics must enter at the TeV scale to cancel quadratic divergences
in the Higgs potential and stabilize the electroweak scale. The most
promising candidate appears to be supersymmetry, with the MSSM being its simplest realization.
If supersymmetry (or any other new physics) is discovered at the LHC 
the analysis performed in the previous section no longer
holds, since the Standard Model will stop to be a good description way before the quartic
Higgs coupling can become negative. In particular in the MSSM the quartic terms
in the Higgs potential are provided by $D$~terms that, being proportional to the
squared gauge couplings, never turn negative. Nevertheless this does not rule
out all the possibilities to test in forthcoming accelerators 
whether the current stage of inflation may be non-eternal. Indeed, the scalar potential of the MSSM includes
also squark fields, and for certain choices of soft terms, non-trivial 
vacua may form. These correspond to non-trivial vev's
for the squarks, which break charge and color.

In this framework the situation is even more interesting than in the case of the Standard Model
since the potential develops, already at tree level, new vacua at the TeV scale, thus their
existence does not rely on assumptions about UV physics. As for the Standard Model, the condition
not to have eternal inflation, together with the requirement to have an enough long-lived
Universe allows only very specific combinations of parameters.
However, the presence of many parameters entering the expression
of the scalar potential makes the actual constraint very model dependent.
In \cite{KLS} the authors studied the tunneling rate for different choices of
the MSSM parameters; they identified an ``empirical region" of the parameter
space with a high probability to encounter a metastable SM-like vacuum, long-lived enough 
to allow the Universe to survive until today. The condition reads
\beq \label{eq:MSSMbound}
3<\frac{A_t^2+3\mu^2}{m_{\tilde t_L}^2+m_{\tilde t_R}^2}<7.5\,,
\eeq
where $A_t$, $\mu$, $m_{\tilde t_{L/R}}$ are the top $A$-term, the $\mu$-term and the stop masses.
At the lower bound \cite{CLM} the Universe becomes metastable, while at
the upper bound  the decay rate starts being faster than the inverse life time of the Universe.
Since the percolation transition occurs when $\Gamma \approx H_\Lambda^4$,
a decay that avoids eternal inflation is expected close to the upper region
of the interval (\ref{eq:MSSMbound}), i.e. when
\beq \label{eq:nodSMSSM}
A_t^2+3\mu^2 \simeq 7.5 (m_{\tilde t_L}^2+m_{\tilde t_R}^2)\,.
\eeq
As before,
this is mainly due  the exponential dependence of the rate on the parameters
of the action and because the window in the decay rate between percolation and the observed 
lifetime of the Universe is very narrow.
Observing MSSM parameters to satisfy relation (\ref{eq:nodSMSSM}) may be interpreted as 
a strong indication that the current acceleration of the Universe is not eternal.
We should say, however, that for particular choices of the MSSM parameters, 
vacuum decays that avoid the eternal inflation phase may happen also
away from the condition (\ref{eq:nodSMSSM}). Indeed, as discussed in \cite{KLS}, 
the bound in eq.~(\ref{eq:MSSMbound}) is 
neither a necessary nor a sufficient condition for metastability.
It is still true however, that the combinations of parameters 
allowing fast enough decays are a minuscule fraction 
of the parameter space so that, measuring the MSSM parameters
in those regions would enforce the case for a non-eternal acceleration of our Universe.

Since non-eternal inflation bounds give very sharp constraints, in order to test
them an equally high precision is needed in the determination of the parameters, 
as discussed in the previous section for the Standard Model.
Indeed the bounce action at the percolation point is expected to be $\sim 400$~\cite{KLS}, (slightly less than
in the SM case since the size of the bubble here is larger) which translates into better than one percent
accuracy in the scalar potential parameters. However, unlike the SM case, 
the parameters of the MSSM are very unlikely
to be measured with the desired precision in the 
forthcoming experiments; still, if the experiments favor such scenario, it is of great
importance to nail down errors as much as we can, to increase (or dump) our confidence on such
interpretation.

Finally we would like to recall that, just as in the Standard Model case, this type 
of bounds
only works in one direction: measuring the parameters in a region that produce a small 
decay rate (with respect to $\eps H_\Lambda^4$), does not rule out the possibility that new physics (at any scale) 
may still trigger new fast decay channels of our vacuum, thus avoiding eternal inflation.

\section{A ``No eternal inflation" principle?}
\label{sec:no}
We see that the  precision measurements of the Higgs mass and other Standard
Model
parameters may provide a crucial information about our future.
If no new physics is found at the LHC and the Higgs mass comes out to be
in the $\sim 0.2$ GeV interval such that our existence now is not very
improbable, but the
percolation transition takes place, this will be a
very strong indication that the current state of the Universe is totally
unstable.
Eventually (and actually quite fast---within the time of order  the current
age of the Universe,
$\sim 10^{10}$ yr) no space-time region able to accommodate life forms even
vaguely resembling ours will exist. By itself,
this would definitely be an interesting and useful thing to know.

However, given that the possibility of the percolation transition requires
a detailed tuning of the Standard Model parameters and there  is no anthropic
reason for the Higgs mass to be below the percolation value, one may wonder
whether this numerical coincidence
would also indicate the existence of some dynamical mechanism operating in the
underlying microscopic theory. Here we outline a line of thinking suggesting
 that underlying fundamental physics must make eternal inflation impossible.

The starting point is that
making sense of quantum gravity in  de Sitter space has proved to be a
notoriously
hard and confusing problem. It is widely believed  that no consistent theory
of pure de Sitter
space is possible (although an alternative viewpoint is also being developed
\cite{Banks:2002nm,Banks:2004xh,Banks:2005bm}).
One piece of evidence
supporting this is related to the impossibility to define the boundary
observables
which can be measured by a single observer in de Sitter space. Recall, that
quantum theory involving dynamical gravity does not allow sharply defined
local observables, so that the only
solid set of data in such a theory is related to the asymptotic quantities,
such as S-matrix elements
in flat space, or boundary CFT correlators in AdS. The non-trivial causal
structure of de Sitter
does not allow to define the appropriate set of boundary observables.

A further piece of evidence comes from the properties of the Coleman--De
Luccia
instanton \cite{Coleman:1980aw}. One can try to realize the eternal de Sitter
space
by starting with a potential with a global Minkowski or AdS minimum and
possessing a local positive energy minimum.  Then,
a region of space filled with the scalar field sitting in the local minimum
is stable under
small fluctuations, but non-perturbative creation of the true vacuum bubbles
causes it to decay.
In field theory one can consider a limit when the height and/or width of the
barrier separating the
two vacua grows indefinitely. In this limit the bubble nucleation rate
vanishes and one smoothly approaches a theory with a stable positive energy
vacuum. With dynamical gravity turned
on, the remarkable property of the Coleman-De Luccia instanton describing this
transition is that
independently of the height of the barrier
its action never drops below the Poincar\'e recurrence time $e^{-S_{\rm dS}}$, where $S_{\rm dS}\sim (M_{Pl}/H)^{2}$
is the entropy of
de Sitter vacuum. Sometime (for instance, in the limit of a very broad
barrier)
the Coleman--De Luccia instanton ceases to exist, but this does not imply that
the instability rate
may drop below $e^{-S_{\rm dS}}$. Instead, in this case the decay goes through the
Hawking--Moss
instanton \cite{Hawking:1981fz}, which physically describes a thermal jump of
the field onto the top of the barrier. The fact that it is impossible to
increase the life-time of metastable de Sitter  phase
indefinitely, strongly suggests that pure de Sitter does not exist.
This conclusion
is further supported by  string theory where  de Sitter phases are always
metastable.

If one adopts the point of view that de Sitter space always corresponds to a
metastable
state, then the conventional logic implies that all physical information about
this state is encoded in the S-matrix elements (boundary correlators)
of the underlying stable Minkowski (AdS) vacuum. More specifically, in the
above field theory example
one may consider a process where, for instance, a large number of soft scalar
quanta in the true vacuum
collide and produce a big bubble of the false vacuum. Later this bubble
shrinks (or, if its size is
really huge, bubbles of the true vacua are produced non-perturbatively within
it) and decays again in a large number of scalar quanta around the true
vacuum. If in addition a number of hard quanta is added
to the scattering process, the corresponding amplitude can be conveniently
approximated
by solving the field equations for the hard quanta in the classical soft
background.
 By considering all possible matrix elements of this sort
one can reconstruct all physical properties of the false vacuum; in fact these
matrix elements
provide the only unambiguous way to define what one understands by ``false
vacuum".

However,  applying this logic in the presence of gravity one again faces a
problem. Indeed, Guth
and Farhi have shown \cite{Farhi:1986ty,Farhi:1989yr} that, at least at the
classical level, it is not possible to start with a non-singular
perturbation around the flat vacuum and create a bubble of the false vacuum
containing more than one
Hubble volume of de Sitter space. Whenever one attempts to create an inflating
bubble
of the false vacuum its interior collapses and the bubble turns into a black
hole instead of creating
exponentially inflating Universe inside.

Nevertheless, one might proceed by assuming that at the quantum level it may
be possible to get around the classical
obstacle preventing to form a bubble of an inflating Universe in a collision
of excitations around
stable vacuum. This was actually done in the original Guth--Farhi paper who
suggested a singular instanton solution describing such a process.
This assumption is not enough, however, to allow the extraction of de Sitter
properties from scattering matrix elements.
Indeed, to achieve this, one also need to end up in an out-state, {\it i.e} the
bubble should again decay into
a set of small perturbations around the true vacuum. However,  if the bubble
interior corresponds to an eternally inflating Universe (as is the case, for
instance, of the Guth--Farhi instanton), the causal structure
at late times  does not resemble that of the flat space even remotely.

A more detailed and quantitative discussion of these issues can be found in
\cite{Freivogel:2005qh},
where the creation of inflating bubbles is  addressed in AdS space. In this
case AdS/CFT allows
to formulate the problem even more sharply, as the creation of an inflating bubble
would correspond to a transition from a pure to a mixed state in the dual
CFT, which is impossible. A somewhat different attempt to construct an
S-matrix description of  de Sitter space using the Lorentzian Coleman--De~Luccia 
solution \cite{Freivogel:2004rd} was also found to be problematic due
to the
rapid instabilities plaguing this solution \cite{Bousso:2005yd}.

All of these issues would be resolved in a straightforward though brutal way if
inflation inside the bubble were not
eternal, {\it i.e.}, if a consistent theory of gravity were not able to
support the metastable vacua
with decay rates below the critical value (or the inflationary potentials
supporting
eternal inflation). Indeed such a possibility has already been proposed by
Page~\cite{Page:2006nt} to avoid another very confusing feature of
eternal inflation---the proliferation of Boltzmann
brains in a global picture of the spacetime.
It well may be that this idea is too radical---it definitely requires
conspiratorial gravitational restrictions on particle physics, although it is
intriguing that the
 bubble size in the SM case is not that far from the Planck scale, so that
gravitational corrections may start being
 important.

Although it seems more plausible that the eternal inflation does make sense
and we have to learn how to deal with it,
at the moment this ``no eternal inflation" proposal  does not
contradict any experimental or observational data.
On the theoretical side, no realization of the slow roll eternal inflation has
been found in string theory so far,
and the available constructions of the metastable de Sitter vacua are  too
approximate to
definitely exclude the possibility of non-perturbative processes giving rise
to faster than
critical decay rates. On the observational side it does not preclude sixty
e-foldings of slow
roll inflation, and as we saw, might acquire a dramatic support from the
future particle physics data.

Let us make a last comment. Eternal inflation is often considered as a
necessary ingredient
for the environmental solution to the cosmological constant problem. Indeed,
without eternal inflation, it
appears problematic to efficiently populate the landscape of vacua. However,
we would like
to stress that it is the mere existence of a vast landscape that
makes it possible to find vacua with un-naturally
small values of the cosmological constant and perhaps other parameters.
How the vacua are realized in nature is a separate issue.
We do not know whether eternal inflation is the only solution; actually,
we are not even sure whether the question of ``population" is well-defined:
it may well be that even with eternal inflation, the naive semiclassical
picture of
how the landscape gets populated is misleading (see e.g.
\cite{ArkaniHamed:2007ky}). The
``no eternal inflation" principle thus neither conflicts nor supports
the landscape
approach
to the cosmological constant problem.

%A last comment we want to make is about the relation with the landscape of
%vacua.
%Eternal inflation is often considered as the only concrete way to populate
%the landscape and thus is also considered as a requirements for

\section{Discussion}
To summarize,  forthcoming particle physics data may provide a surprising
twist for fundamental physics and cosmology,
even in the ``nightmare scenario" where nothing but a light Higgs is
observed at the LHC.
As has long been known, our vacuum may be metastable for a light enough Higgs.
For a very narrow window of parameters, our Universe has not yet decayed
but the current inflationary period can not be future eternal.
To support this conclusion one would need to measure the Standard Model
parameters
with extreme precision, requiring significant but achievable progress
in the theoretical calculations and
experimental measurements.

As a benchmark value for the desired accuracy  we used
$0.2$~GeV for the Higgs mass. Given the relation (\ref{eq:res}) this
corresponds to the precision
at the level of 60~MeV for the top mass (as opposed to the present $\sim
1.8$~GeV) , and
at the level of $0.14$\% for the strong coupling (as opposed to the current
precision
at the level $\sim1.7$\%). With this precision one would be able to
conclusively establish that the current
stage of the cosmological acceleration is not eternal if the decay rate of our
vacuum is fast enough, so that
the probability that we did not decay up to now is $5$\%. Even higher
precision is needed if
the decay rate is slower. Note that  in this case, even if we are not able
to resolve whether the Higgs mass lies above or below the percolation bound,
finding  the Higgs within
$0.2$~GeV from that bound would be remarkable enough to take seriously the
idea that eternal inflation
may be impossible. This may be true for the heavier Higgs masses, when the
incalculable  gravitational corrections may in principle affect the bound at
this level of precision.

 For the Higgs mass itself the desired level of accuracy is likely to be
achievable  directly at the LHC
 \cite{TDRs}. Indeed, for a light  Higgs, a very precise determination of the
mass is possible
 by measuring the location of the maximum of the photon distribution from the
$H\to\gamma\gamma$
 decays. It is important that we need the mass itself;
 much higher statistics would be needed to determine the width with the same
precision.

 The top mass precision will not improve much at the
LHC~\cite{TDRs,Stavropoulos:2005he}. The statistical uncertainties will be
reduced quite significantly and will be practically negligible.
The problem is due to the large systematic
 uncertainties from the jet energy scale and final state radiation. The best
one can hope for is
 $\Delta m_t\sim1$~GeV. However, the situation improves a lot at a linear
collider \cite{Hoang:2000ib,Penin:2002zv}. Here one can use
 the shape of the cross-section for $t\bar{t}$ production near threshold to
determine the top mass.
 This method is analogous to the extraction of the bottom mass from the
spectrum of the bottomonium.
 At the moment the theoretical uncertainty from this method has been reduced
to the level $\sim 80$~MeV and
 may  improve somewhat in the future.

 The best determination of $\alpha_s$ at the moment (and apparently in the
reasonable future as well)
 is from lattice QCD. It appears reasonable  to expect an order of magnitude
improvement
 ({\it i.e.}, the required level) within the time-scale of constructing a new
linear collider.

 Finally, on the theoretical side one will need to improve both the running
and the matching conditions
 for gauge couplings and masses by at least one extra electroweak loop and two
strong loops.
This also requires to include another loop for the calculation of the bounce
action.

It is worth mentioning that there is a theoretically motivated natural level
of precision, that one may wish to achieve.
Indeed, even if the Universe is eternally inflating,  bubbles of new vacuum
may
form infinite volume clusters if the decay rate is fast enough. The exact
value of $\eps\equiv\eps_p$ when this
transition (usually referred to as the ``percolation transition", contrary the
terminology adopted in the current paper) happens is currently unknown and is
limited to be in the range (\ref{truepercolation}).
Most likely $\eps_p$ is quite close to the upper boundary $\eps_p\sim
0.24$. It appears natural to try to achieve
an accuracy allowing to distinguish this critical value $\eps_p$ from
$\eps=9/4\pi$
 (corresponding to the transition to the
eternal regime). If $\eps_p=0.24$ this would require the Higgs mass
resolution at the level of
$0.05$~GeV, but may be somewhat less challenging if $\eps_p$ is smaller.

There are
other scenarios motivating the Higgs mass to be in the proximity of the
life-time bound.
For instance, a  proposal of Ref.~\cite{Feldstein:2006ce} is based on specific
assumptions on how the
Standard Model parameters scan in the landscape. More generally, this may be a
natural expectation
if one adopts the ``living dangerously"
logic~\cite{Weinberg:1987dv,ArkaniHamed:2005yv}.

The key observation of the current paper is the existence of a sharp critical
value
of the Higgs mass,  having a direct physical meaning independently of any
assumptions on the landscape statistics and on what one understands by being
``close" to the life-time bound.
A high precision
is of crucial importance to take the full advantage of the existence of this
sharp number.
One of the benefits of having a sharp prediction, is that although we will
never be able
to directly test that there is no new physics up to the GUT scale, finding the
Higgs mass in the
correct window would provide a strong case for such a scenario.There
are also very well-motivated ideas for physics at the TeV scale, such
as low-energy SUSY with large $A$~terms, where we can conclude that our
vacuum is metastable. In this case high-precision measurements can in
principle again tell us that the instability rate is subcritical for
eternal inflation, and in this case the conclusion can be drawn much
more sharply since the physics of the instability involves no further
theoretical extrapolation and can be studied at the TeV scale.

\section*{Acknowledgments}We would like to thank Paolo Creminelli, Lawrence Hall, Beate Heinemann, Gino Isidori, Sasha Penin and Matias Zaldarriaga for useful discussions. Our work is supported by the DOE under contract DE-FG02-91ER40654.

\section*{Appendix}
In this Appendix we give some details of the calculation of the decay rate of the Standard Model vacuum, and  provide the most relevant formul\ae. 

We need only the Higgs quartic $\lambda$, the gauge $g$, $g'$, $g_3$, and top Yukawa $h$ couplings, and work in the $\overline{\rm MS}$ scheme. The normalization of the Higgs quartic coupling is chosen so that the tree-level potential for the physical Higgs $\phi_h$ reads
\begin{equation}
V(\phi_h)=\frac{1}{24}\lambda\l(\phi_h^2-v^2\r)^2 \ ,
\end{equation}
where $v=(\sqrt2 G_\mu)^{-1/2}=246.221$~GeV and $G_\mu=1.16637\cdot 10^{-5}$~GeV$^{-2}$ is the Fermi constant from muon decays.
As explained in the main part of the text, we solve the RG equations for the running couplings to
${\cal{O}}(\alpha_s^3)$, ${\cal{O}}(\alpha_{W}^2)$, ${\cal{O}}(\alpha_s \alpha_{W})$, which means two-loop plus pure $\alpha_s$ three-loop. For consistency, we match the initial conditions for the top Yukawa and the Higgs quartic coupling up to order ${\cal{O}}(\alpha_{W})$ (one loop), and ${\cal{O}}(\alpha_s^2)$ (pure two loops in $\alpha_s$). Higher order corrections to the $\beta$-functions and to the matching conditions (four loops in $\alpha_s$ to the strong coupling running, three loops in $\alpha_s$ and two loops mixed strong/weak to the top Yukawa matching) are known. These form an incomplete list of the quantities required for doing a consistent next order computation, and therefore we use them to estimate the theoretical errors.
In order to reduce the theoretical uncertainty associated with the available expressions for the matching conditions, we perform the matching respectively for the top Yukawa at the top mass scale, for the Higgs quartic at the Higgs mass scale, and for the gauge couplings at the $m_Z$ scale.  

We are now ready to give the expression for the relevant equations.
The RG equations are given by
\begin{eqnarray}
\frac{d}{d t} \lambda(t)&=&\kappa \beta_\lambda^{(1)}+\kappa^2  \beta_\lambda^{(2)}\,, \\  \nonumber
\frac{d}{d t}h(t)&=&\kappa\beta_{h}^{(1)}+ \kappa^2  \beta_{h}^{(2)}+ \kappa^3 \beta_{h}^{(3)}\,,\\  \nonumber
\frac{d}{dt} g(t)&=&\kappa\beta_g^{(1)}+\kappa^2   \beta_{g}^{(2)}\,,\\  \nonumber
\frac{d}{dt} g'(t)&=&\kappa\beta_{g'}^{(1)}+ \kappa^2  \beta_{g'}^{(2)}\,,\\  \nonumber
\frac{d}{dt} g_3(t)&=&\kappa\beta_{g_3}^{(1)}+ \kappa^2  \beta_{g_3}^{(2)}+ \kappa^3 \beta_{g_3}^{(3)}
+ \kappa^4 \beta_{g_3}^{(4)}\,, \nonumber
\end{eqnarray}
where $t=\log(\mu/m_Z)$ with $\mu$ being the renormalization scale. 
The apex on the $\beta$-functions represents the loop order. They are given by \cite{FJSE,T,L2,TVZ,LV,vRVL,Czakon:2004bu}:
\begin{eqnarray}\nonumber
\beta_\lambda^{(1)}&=&  \frac{27}{4} g(t)^4+\frac{9}{2} g'(t)^2 g(t)^2-9 \lambda (t) g(t)^2+\frac94
   g'(t)^4-36 h(t)^4+4 \lambda (t)^2-3 g'(t)^2 \lambda (t) \\ && +12 h(t)^2 \lambda (t)\ , \\  \nonumber
\beta^{(1)}_h&=& \frac92 h(t)^3-\frac{9}{4} g(t)^2 h(t)-8
g_3(t)^2 h(t)-\frac{17}{12} g'(t)^2
  h(t)\ , \\ \nonumber
\beta_g^{(1)}&=&-\frac{19}{6}   g(t)^3\ ,\\  \nonumber
\beta_{g'}^{(1)}&=&\frac{41}{6}   g'(t)^3\ ,\\ \nonumber
\beta_{g_3}^{(1)}&=&-7  g_3(t)^3\ ,\\
%\end{eqnarray}
%\begin{eqnarray}
\nonumber \beta_\lambda^{(2)}&=&80g_3(t)^2 h(t)^2 \lambda (t)-192g_3(t)^2 h(t)^4+ 
   \frac{915}{8} g(t)^6-\frac{289}{8} g'(t)^2 g(t)^4-\frac{27}{2} h(t)^2 g(t)^4\\  \nonumber &&  
   -\frac{73}{8} \lambda (t) g(t)^4-\frac{559}{8} g'(t)^4 g(t)^2+63 g'(t)^2 h(t)^2 g(t)^2+\frac{39}{4} g'(t)^2 \lambda (t)
   g(t)^2-3 h(t)^4 \lambda (t)\\ &&  \nonumber +\frac{45}{2} h(t)^2 \lambda (t) g(t)^2
   -\frac{379}{8} g'(t)^6+180 h(t)^6-16 g'(t)^2 h(t)^4-\frac{26}{3}\lambda (t)^3
   -\frac{57}{2} g'(t)^4 h(t)^2\\ &&  \nonumber -24 h(t)^2 \lambda (t)^2   +6 \left(3 g(t)^2+g'(t)^2\right) \lambda
   (t)^2+\frac{629}{24} g'(t)^4 \lambda (t)+\frac{85}{6} g'(t)^2 h(t)^2 \lambda (t)
   \ , \\ \nonumber
\beta_{h}^{(2)}&=&h(t) \left[-108g_3(t)^4+9 g(t)^2g_3(t)^2+\frac{19}{9} g'(t)^2g_3(t)^2+36 h(t)^2
  g_3(t)^2-\frac{3}{4} g'(t)^2 g(t)^2\right. \\ && \nonumber
  \left. -\frac{23}{4} g(t)^4+\frac{1187}{216}g'(t)^4-12 h(t)^4+\frac{\lambda (t)^2}{6}+h(t)^2 \left(\frac{225}{16} g(t)^2+\frac{131}{16} g'(t)^2-2
   \lambda (t)\right)\right] \ ,\\ \nonumber
\beta_{g}^{(2)}&=&12 g_3(t)^2 g(t)^3+   \left(\frac{35}{6} g(t)^2+\frac32 g'(t)^2-\frac32 h(t)^2\right) g(t)^3\ ,\\ \nonumber
\beta_{g'}^{(2)}&=&\frac{44}{3} g_3(t)^2 g'(t)^3+   \left(\frac92 g(t)^2+\frac{199}{18} g'(t)^2-\frac{17}{6} h(t)^2\right) g'(t)^3\ ,\\ \nonumber
\beta_{g_3}^{(2)}&=&g_3(t)^3 \left(\frac92 g(t)^2-26g_3(t)^2+\frac{11}{6} g'(t)^2-2 h(t)^2\right)\ ,\\ \nonumber
   \beta_{h}^{(3)}&=&384 g_3(t)^6 \left(-\frac{2083}{576}+\frac53\zeta_3\right)\ ,\\ \nonumber
\beta_{g_3}^{(3)}&=&\frac{65}{2} g_3(t)^7\ ,\\ \nonumber
\beta_{g_3}^{(4)}&=&g_3(t)^9 \left ( \frac{63559}{18} -\frac{44948}{9}\zeta_3 \right ) \ .
\end{eqnarray}
where $\zeta_3=1.20206...$ is the Riemann zeta function and
\begin{equation}
\kappa\equiv\frac{1}{16 \pi ^2} \ .
\end{equation}
Notice that the 4-loop QCD correction to the $\beta$-function of $g_3$ ($\beta_{g_3}^{(4)}$) is an higher order effect
and has only been used to estimate the theoretical uncertainty.
The matching between the top pole mass and the $\overline{\textrm{MS}}$ Yukawa is given by
\begin{equation}
h(\mu)=2^{3/4} \sqrt{G_\mu} m_t \Bigl(1+  \delta_t(\mu)\Bigr)\ ,
\end{equation}
where 
\begin{eqnarray}
\delta_t(\mu)=\delta_t^{QCD}(\mu)+\delta_t^{W}(\mu)+\delta_t^{QED}(\mu)\ .
\end{eqnarray}
Here $\delta_t^W+ \delta_t^{QED}$ represent the one loop electroweak contribution and is given by \cite{HK}
\begin{eqnarray}
&&\delta_t^W(\mu)+\delta^{QED}_t(\mu)=-\frac{E(\mu)}{2}+{\rm{Re}}\Bigl[\Sigma_V(\mu)+\Sigma_S(\mu)\Bigr]-\frac{\Pi (\mu)}{2 m_W^2} \ ,
\end{eqnarray}
where
\begin{eqnarray}
   E(\mu)&=&\frac{\alpha_{em}(m_Z) }{4\pi s^2}\left[\left(\frac{7}{2 s^2}-6\right) \log \left(c^2\right)-4 \log
   \left(\frac{m_Z^2}{\mu^2}\right)+6\right] \ , \\
\Sigma_V(\mu)+\Sigma_S(\mu)&=&\frac{\alpha_{em}(m_Z)}{4\pi} \left\{\left(6-\frac{m_Z^2}{m_t^2}\right) a_f^2-4
   Q_t^2-\left(\frac{m_Z^2}{m_t^2}+4\right) v_f^2 \right. \\  \nonumber && \left. 
   +\left[a_f^2
   \left(4-\frac{m_Z^2}{m_t^2}\right)-\left(\frac{m_Z^2}{m_t^2}+2\right) v_f^2\right]
   F\left(m_t^2,m_t^2,m_Z^2\right) \right. \\ && \left.\nonumber
   -\left[\frac{3}{8 s^2}
   \left(\frac{m_t^2-m_b^2}{m_W^2}+1\right)-3
   \left(Q_t^2+v_f^2-a_f^2\right)+\frac{1}{8 c^2}\right] \log
   \left(\frac{m_t^2}{\mu^2}\right)\right. \\ && \left. \nonumber
   +\frac{1}{4 m_W^2 s^2}\left[ 4 m_t^2-\frac{5 m_b^2}{2}+\frac{1}{2}
   \left(m_W^2-m_H^2\right)+\frac{m_b^4+m_b^2 m_W^2-2 m_W^4}{2 m_t^2}\right. \right. \\ && \nonumber
   \left. \left. +\frac{1}{2 m_t^2}\left(\left(m_b^2+m_t^2\right) m_W^2+\left(m_t^2-m_b^2\right)^2-2 m_W^4\right)
   F\left(m_t^2,m_b^2,m_W^2\right)\right.\right. \\ && \nonumber  \left. \left. +\left(2 m_t^2-\frac{m_H^2}{2}\right)
   F\left(m_t^2,m_t^2,m_H^2\right)+\left(m_W^2+\frac{1}{2} \left(m_t^2-3
   m_b^2\right)\right) \log \left(\frac{m_t^2}{m_b^2}\right)\right. \right.\\ && \nonumber
   \left.\left. +\frac{1}{2} m_H^2
   \left(3-\frac{m_H^2}{2 m_t^2}\right) \log \left(\frac{m_t^2}{m_H^2}\right)
   +\frac{1}{4 m_t^4}\Bigl (3 \left(m_b^2+m_t^2\right) m_W^4+4 m_t^4
   m_W^2\right. \right. \\ && \left.\left.\nonumber
    +\left(m_t^2-m_b^2\right)^3-2 m_W^6\Bigr) \log \left(\frac{m_b^2}{m_W^2}\right)\right]\right.
	\\ && \nonumber \left.
	+\left[a_f^2 \left(2-\frac{m_Z^4}{2 m_t^4}+\frac{3
   m_Z^2}{m_t^2}\right)-\frac{m_Z^4 v_f^2}{2 m_t^4}\right] \log
   \left(\frac{m_t^2}{m_Z^2}\right)\right\}\ , \\ 
\Pi (\mu)&=&\frac{\alpha_{em}(m_Z) m_W^2}{4\pi s^2} \left[\frac{7}{8 c^2}-\frac{17}{4}-\frac{3 m_H^2}{4
   \left(m_W^2-m_H^2\right)} \log \left(\frac{m_W^2}{m_H^2}\right)-\frac{m_H^2}{8 m_W^2}
   \right.\\ \nonumber && \nonumber \left.
   +\left(2+\frac{1}{c^2}-\frac{17}{4 s^2}\right) \log
   \left(c^2\right)-\left(\frac{1}{c^2}-2\right) \log \left(\frac{m_W^2}{\mu^2}\right)\right]\\ 
 && \nonumber  +\frac{3 \alpha_{em}(m_Z) }{4\pi s^2}\left[ \frac{m_b^2 m_t^2 }{m_t^2-m_b^2}\log\left(\frac{m_t^2}{m_b^2}\right)
 -\left(\frac{1}{2}-\log\left(\frac{m_b^2}{\mu^2}\right)\right) m_b^2  \right. \\ && \nonumber \left.
  -m_t^2 \left(\frac{1}{2}-\log
   \left(\frac{m_t^2}{\mu^2}\right)\right)\right] \ ,
\end{eqnarray}
with
\begin{equation}
    F(x,y,z)=\left\{ \begin{array}{ll} \displaystyle
\frac{\tilde\lambda (x,y,z)^{\frac12}}{x} \, {\rm arccosh}\left(\frac{-x+y+z}{2 \sqrt{y z}}\right)  & \textrm{if}\ x<  \left(\sqrt{y}-\sqrt{z}\right)^2 \\  &\\ 
\displaystyle
 -\frac{\sqrt{-\tilde\lambda (x,y,z)}}{x}\,  {\rm arccos}\left(\frac{-x+y+z}{2 \sqrt{y z}}\right) & \textrm{if}\ \left(\sqrt{y}-\sqrt{z}\right)^2  \leq x\leq  \left(\sqrt{y}+\sqrt{z}\right)^2  \\  &\\ 
   \displaystyle \frac{\tilde\lambda (x,y,z)^{\frac12} }{x}\left[i \pi -{\rm arccosh}\left(\frac{x-y-z}{2 \sqrt{y z}}\right)\right]  & \textrm{if}\ x>  \left(\sqrt{y}+\sqrt{z}\right)^2\ ,
   \end{array}
   \right.     \
\end{equation}
and
   \begin{equation}
\tilde\lambda (x,y,z)=x^2+y^2+z^2-2 (x
   y+z y+x z) \ ,
   \end{equation}
%and
\begin{equation}
s=\sqrt{1-\frac{m_W^2}{m_Z^2}}, \quad \quad
c=\frac{m_W}{m_Z}, \quad Q_t=\frac{2}{3},\quad \quad
v_f=\frac{1}{4sc}-\frac{2 s}{3c}, \quad \quad
a_f=\frac{1}{4s c}\ . 
\end{equation}
The numerical values we used are~\cite{PDG}
\begin{eqnarray}
m_b=4.2~\textrm{GeV},  \quad m_Z=91.1876~\textrm{GeV},  \quad
m_W=80.403~\textrm{GeV}, \quad
\alpha_{em}^{-1}(m_Z)=127.9\,.
\end{eqnarray}
Concerning the pure QCD contribution to $\delta_t$, we have the one, two and three loop results ~\cite{GBGS,BGS,FJTV} 
\begin{eqnarray}
&&
\delta^{{\rm QCD},\,(1)}_t(\mu)=-\frac{4 \alpha_s(\mu)}{3\pi} \left(1-\frac{3}{4} \log
   \left(\frac{m_t^2}{\mu^2}\right)\right)\ ,\\&&
\delta^{{\rm QCD},\,(2)}_t(m_t)=(-14.3323+1.0414\times 5 )\left(\frac{\alpha_s(m_t)}{\pi}\right)^2\ ,\\&&
\delta^{{\rm QCD},\,(3)}_t(m_t)=(- 0.65269\times 5^2+26.9239 \times 5-198.7068)\left(\frac{\alpha_s(m_t)}{\pi}\right)^3\ . \label{eq:3-loop top matching}
\end{eqnarray}
Notice that the pure 2-loop and 3-loop contributions  are known only at the top mass scale, and this is why we perform the matching for the top at this scale. We do not use $\delta^{{\rm QCD},\,(3)}_t$ directly in deriving our bound (\ref{eq:res}), but just to estimate the theoretical uncertainty. Further, 
the mixed 2-loop strong/weak matching correction is known~\cite{JK}, and it is numerically comparable
to the effect of (\ref{eq:3-loop top matching}). 

Turning to the Higgs, we match the Higgs pole mass to the $\overline{\rm {MS}}$ quartic coupling at one loop through the following~\cite{SZ}
\begin{equation}
\lambda(\mu)=3 \sqrt{2} G_\mu m_H^2 \Bigl(1+ \delta_H(\mu)\Bigr)\ ,
\end{equation}
where
\begin{equation}
\delta_H (\mu)=\frac{G_\mu m_Z^2}{8 \sqrt{2} \pi ^2} \left(\xi  f_1(\mu)+f_0(\mu)+\frac{f_{-1}(\mu)}{\xi
   }\right) \ ,
\end{equation}
with
\begin{eqnarray}
\xi &=&\frac{m_H^2}{m_Z^2}\ ,\\
f_1(\mu)&=&\frac{3}{2} \log (\xi )-\log \left(c^2\right)+6 \log \left(\frac{\mu^2}{m_H^2}\right)-\frac{1}{2}
   Z\left[\frac{1}{\xi }\right]-Z\left[\frac{c^2}{\xi }\right]+\frac{9}{2} \left(\frac{25}{9}-\frac{\pi }{\sqrt{3}}\right)\,,\\
f_0(\mu)&=&\frac{3  c^2}{s^2}\log \left(c^2\right)+12 \log c^2\left(c^2\right) +\frac{3 \xi   c^2}{\xi -c^2}\log \left(\frac{\xi
   }{c^2}\right)+4c^2\,Z\left[\frac{c^2}{\xi }\right] -\frac{15}{2} \left(2 c^2+1\right)\\
   \nonumber && -6 \left(2 c^2-\frac{2m_t^2}{m_Z^2}+1\right) \log \left(\frac{\mu^2}{m_Z^2}\right)
   -\frac{3m_t^2}{m_Z^2} \left(4 \log \left(\frac{m_t^2}{m_Z^2}\right)
   +2\,Z\left[\frac{m_t^2}{m_Z^2 \xi}\right]-5\right)\\
   &&+2\, Z\left[\frac{1}{\xi }\right]\,, \\ \nonumber 
f_{-1}(\mu)&=&8 \left(2 c^4+1\right)-12c^4 \log \left(c^2\right) -12c^4\, Z\left[\frac{c^2}{\xi }\right] +6 \left(2
   c^4-\frac{4 m_t^4}{m_Z^4}+1\right) \log \left(\frac{\mu^2}{m_Z^2}\right)\\
   &&-6\,Z\left[\frac{1}{\xi }\right]+\frac{24 m_t^4 }{m_Z^4}\left(\log \left(\frac{m_t^2}{m_Z^2}\right)+Z\,\left[\frac{m_t^2}{m_Z^2
   \xi }\right]-2\right)\,,\\
Z[z]&=&\left\{ \begin{array}{lll}
2 A(z) \,{\rm arctan}\left(\frac{1}{A(z)}\right) &  \textrm{if} & z>\frac{1}{4} \\
A(z) \log \left(\frac{A(z)+1}{1-A(z)}\right) &  \textrm{if} & z<\frac{1}{4}
\end{array}
\right. \,, \\
A(z)&=&\sqrt{|1-4 z|} \,.
\end{eqnarray}

We finally give the formula for $\Delta S$~\cite{IRS}
\begin{eqnarray}\nonumber
\Delta S(t)&=&6 \l(6 L(t)+5\r)\frac{ h(t)^4}{\lambda(t)^2}+\l(12 L(t)+13\r)\frac{h(t)^2}{|\lambda(t)|}
-\frac{2}{3} \l(6 L(t)+5\r)\\  \nonumber
&& -\l(6 L(t)+7\r) \frac{2  g(t)^2+g_Z(t)^2}{2 |\lambda(t)|}
-9 \l(2 L(t)+1\r) \frac{ 2   g(t)^4+g_Z(t)^4}{8 \lambda(t)^2}\\ \nonumber &&
+f_h\l(\lambda(t)\r)+2 f_g\left(\frac{6 g(t)^2}{|\lambda
   (t)|}\right)+f_g\left(\frac{6 g_Z(t)^2}{|\lambda(t)|}\right)-f_t\left(\frac{6 h(t)^2}{|\lambda(t)|}\right)\ , \\
\end{eqnarray}
where 
\begin{equation}
L(t)=\log\left(\frac{R\; m_Z\;  e^{t+\gamma_E}}{2}\right), \quad \quad \quad g^2_Z=g^2+g'^2\ .
\end{equation}
Here $R$ is the radius of the bubble, and $f_t,\ f_g,\ f_h$ are functions whose numerical values can be found plotted in~\cite{IRS}\footnote{We acknowledge the authors of \cite{IRS} for providing us unpublished numerical
fits of these functions.}. In the numerical results given in the text the renormalization scale has
been set to $\mu=2e^{-\gamma_E}/R$, so that $L(t)=0$; the dependence on the choice of the scale $\mu$ 
has been checked to be small.

\end{document}